\documentclass[onecolumn,10pt]{IEEEtran}

\usepackage{amsfonts,amsmath,amssymb,amsthm,caption,algorithm,algorithmic}
\usepackage{amsbsy}
\usepackage{graphicx,multirow,bm,array,tabularx}
\usepackage{color}
\makeatletter
\newtheoremstyle{mythm}{3pt}{3pt}{}{16pt}{\bfseries}{:}{.5em}{}
\theoremstyle{mythm}
\newtheorem{theorem}{Theorem}[section]
\newtheorem{definition}{Definition}[section]

\newtheorem{coro}{Corollary}[section]
\newtheorem{lemma}{Lemma}[section]

\newcommand{\pf}{\noindent{\quad\ \ \bf Proof:}\ }
\newcommand{\INPUT}{\item[\algorithmicrequire]}
\renewcommand{\algorithmicrequire}{\textbf{Input:}}

\begin{document}

\title{Bounds on Traceability Schemes
\author{Yujie Gu and Ying Miao}
\thanks{Y. Gu is with the Department of Policy and Planning Sciences, Graduate School of Systems and Information Engineering, University of Tsukuba, Tsukuba, Ibaraki 305-8573, Japan (e-mail: s1530147@u.tsukuba.ac.jp).}
\thanks{Y. Miao is with the Faculty of Engineering, Information and Systems, University of Tsukuba, Tsukuba, Ibaraki 305-8573, Japan (e-mail: miao@sk.tsukuba.ac.jp).}
}

\maketitle

\begin{abstract}
The Stinson-Wei traceability scheme (known as traceability scheme) was proposed for broadcast encryption as a generalization of the Chor-Fiat-Naor traceability scheme (known as traceability code). Cover-free family was introduced by Kautz and Singleton in the context of binary superimposed code. In this paper, we find a new relationship between a traceability scheme and a cover-free family, which strengthens the anti-collusion strength from $t$ to $t^2$, that is, a $t$-traceability scheme is a $t^2$-cover-free family. Based on this interesting discovery, we derive new upper bounds for traceability schemes. By using combinatorial structures, we construct several infinite families  of optimal traceability schemes which attain our new upper bounds. We also provide a constructive lower bound for traceability schemes, the size of which has the same order with our general upper bound. Meanwhile, we consider parent-identifying set system, an anti-collusion key-distributing scheme requiring weaker conditions than traceability scheme but stronger conditions than cover-free family. A new upper bound is also given for parent-identifying set systems.
\end{abstract}


\section{Introduction}
In 1994, Chor, Fiat and Naor introduced a traitor tracing scheme, the Chor-Fiat-Naor traceability scheme, applied to the broadcast encryption \cite{Chor1994,Chor2000}. To prevent unauthorized users from accessing the data, the data supplier encrypts the data blocks with session keys and gives the authorized users personal keys to decrypt them. Some unauthorized users (pirate users) might obtain some decryption keys from a group of authorized users (traitors). Then the pirate users can decrypt data that they are not entitled to \cite{Stinson1998}. If a pirate decoder is confiscated, the Chor-Fiat-Naor traceability scheme can trace back to at least one traitor, by comparing the number of common base keys between the pirate decoder and each user's personal key, on the assumption that the number of traitors in the collusion does not exceed a predetermined threshold $t$. In 1998, Hollmann, van Lint, Linnartz, and Tolhuizen proposed a digital fingerprinting scheme, based on codes with the identifiable parent property (IPP codes), to protect against piracy of software by embedding fingerprints into the copyrighted contents \cite{Hollmann1998}. Given an IPP code it is possible for every pirate copy (descendant) of digital contents to identify at least one of its parents, that is, those authorized users each assigned with a fingerprint that contribute to the pirate copy, by computing the intersection of all groups of possible parents who can produce the pirate copy, again on the assumption that the number of parents in the collusion does not exceed a predetermined threshold $t$. Both schemes have been extensively studied in the literature, see \cite{Alon2003}--\cite{Blackburn2003IPP,Blackburn2010,Chor1994,Chor2000,Fernandez2010,Jin2007,Kabatiansky2004,LLS2002,Silverberg2003,Staddon2001}, for example. As a matter of fact, although with different scenarios of security protection, these two schemes are essentially the same, except for the requirement on the tracing efficiency.

In 1998, Stinson and Wei generalized the Chor-Fiat-Naor traceability scheme to the Stinson-Wei traceability scheme. As stated in \cite{Stinson1998}, in a broadcast encryption system, the data supplier generates a base set $\mathcal{X}$ of $v$ keys and assigns $w$ base keys to each authorized user, as the user's personal key. All authorized users can recover the session keys $K$, which are used to decrypt the data blocks, by using their personal keys. In the Chor-Fiat-Naor traceability scheme, the set $\mathcal{X}$ of base keys is partitioned into $w$ subsets $\mathcal{S}_1,\ldots,\mathcal{S}_w$ (each of size $v/w$). Each personal key is a transversal of $(\mathcal{S}_1,\ldots,\mathcal{S}_w)$ (i.e., it contains exactly one base key from each $\mathcal{S}_i$). In this case, the pirate decoder generated by several traitors is also a transversal of $(\mathcal{S}_1,\ldots,\mathcal{S}_w)$, since otherwise the pirate decoder can not work. However, in the Stinson-Wei traceability scheme, each personal key is not necessarily a transversal. A personal key can be made up of any selection of $w$ base keys from the set $\mathcal{X}$. The data supplier can use a $w$ out of $v$ threshold secret sharing scheme (such as the Shamir threshold scheme \cite{Shamir1979}, for example) to construct $v$ shares of the key $K$ and then encrypt each share with a base key in $\mathcal{X}$. The pirate decoder can be made up of any $w$ different base keys from the union of each traitor's personal key. If such a pirate decoder is captured and the size of the coalition does not exceed a predetermined threshold $t$, the Stinson-Wei traceability scheme also can reveal at least one traitor in the collusion by detecting the users who share the maximum base keys with the pirate decoder.
In 2009, Collins \cite{Collins2009} suggested parent-identifying set systems (IPP set systems, or IPPSs) for broadcast encryption, which generalize IPP codes. The point of generalization from an IPP code to an IPPS is essentially the same as that from the Chor-Fiat-Naor traceability scheme to the Stinson-Wei traceability scheme, that is, instead of considering $w$-tuples, we consider $w$-subsets. Just as in the case of IPP codes, when a pirate copy is confiscated, the traitor tracing algorithm based on IPPS also needs to compute the intersection of all groups of possible parents with size at most $t$. Compared to the Stinson-Wei traceability scheme, the traitor tracing scheme based on IPPS can accommodate more users, but at the expenses of tracing efficiency.

The Chor-Fiat-Naor traceability scheme is popular with the notion of ``traceability code (TA code)", which has been studied in   \cite{Blackburn2010,Chor1994,Chor2000,Jin2007,Kabatiansky2004,LLS2002,Staddon2001}, and the Stinson-Wei traceability scheme has been studied as ``traceability scheme (TS)" in  \cite{Blackburn2003,Collins2009,LL2010,Naini2001,Stinson1998,Wei1998}, for example. Objects related with traceability schemes, such as key distribution patterns, also have been studied by numerous researchers, see \cite{Staddon1997,Stinson2000,Wei1998}.
In this paper, we will focus on $t$-TS and $t$-IPPS, both of which can resist the collusion attack with at most $t$ traitors. We call $t$ the \textit{strength} of the scheme.

Cover-free families (CFFs) were introduced in 1964 by Kautz and Singleton \cite{Kautz1964} to study binary superimposed codes. Variants of this formulation have been investigated related to subjects such as information theory, group testing, combinatorics, see \cite{Dyachkov1982}--\cite{Erdos1985,Furedi1996,Hwang1987}, for example. A $t$-CFF is a family of finite sets (blocks) in which no block is covered by the union of $t$ others. From the point of broadcast system, a $t$-CFF is a kind of scheme in which any $t$ traitors can not create another authorized user's personal key, which is closely related to the frameproof code. Frameproof codes were studied by numerous researchers, see \cite{Blackburn2003FP,Boneh1998,Chee2012,Staddon2001,Stinson1998}, for example.

Among the known results on $t$-TS, Stinson and Wei \cite{Stinson1998} proved that a $t$-TS is a $t$-CFF, and derived an upper bound for the number of blocks in a $t$-TS by using this relationship. There is a huge gap between this upper bound and the size of $t$-TS constructed by using combinatorial structures in \cite{Stinson1998}. In \cite{Naini2001}, Safavi-Naini and Wang used constant weight codes to derive a lower bound for $t$-TS. However, L\"{o}fvenberg and Larsson \cite{LL2010} pointed out that there were mistakes in deriving the lower bound for $t$-TS in \cite{Naini2001}. Collins \cite{Collins2009} improved the upper bound for $t$-TS, and gave upper bounds for $t$-IPPS. However, there is no construction which can achieve any of these known upper bounds. As a matter of fact, the known upper bounds for $t$-TS and $t$-IPPS are not tight. In this paper, we will provide new upper bounds for $t$-TS and $t$-IPPS, which greatly improve the previously known upper bounds. Moreover, we will give some constructions which can produce infinite families of $t$-TS achieving our new upper bounds.

As far as we know, in the literature, the relationship between TS and CFF has been studied for the same strength (i.e. a $t$-TS is a $t$-CFF), and this is also almost true for other relationships among various anti-collusion schemes. In this paper, we find a very interesting phenomenon, that is, a $t$-TS is in fact a $t^2$-CFF. This is the first relationship between two kinds of anti-collusion schemes which strengthens the strength from $t$ to $t^2$. Based on this important discovery, new upper bounds for $t$-TS are derived. To obtain our new bounds, we use a combinatorial structure called  \textit{own-subset} by Erd\"{o}s, Frankl, and F\"{u}redi in \cite{Erdos1985}. In a $t$-TS, we show that the number of $\tau$-own-subsets of each block is at least $\binom{w-1}{\tau-1}$, where $\tau=\lceil {w}/{t^2}\rceil$. By applying the double-counting method, we derive our upper bound for general $t$-TS. When $w\le t^2$, we provide a construction for $t$-TS which achieves our general upper bound. Furthermore, we provide a better upper bound for several special cases, which shows that some infinite families of $t$-TSs constructed by Stinson and Wei \cite{Stinson1998} from combinatorial designs are in fact optimal. We generalize Stinson and Wei's construction to obtain more infinite families of optimal $t$-TS for $w>t^2$. We also describe a constructive lower bound for general $t$-TS, the size of which has the same order with our general upper bound. Collins \cite{Collins2009} showed that the upper bound for $t$-IPPS is $O(v^{\lceil\frac{w}{\lfloor t^2/4\rfloor+\lceil t/2\rceil}\rceil})$. We give an improvement for this by showing that the upper bound for $t$-IPPS is $O(v^{\lceil\frac{w}{\lfloor t^2/4\rfloor+t}\rceil})$, which is realized by analyzing the minimum size of own-subsets possessed by some blocks in a $t$-IPPS.

The paper is organized as follows. Firstly, we recap some definitions and notations in Section II. In Section III, we show a new upper bound for $t$-IPPS. New relationship between TS and CFF, and new upper bounds for $t$-TS are provided in Section IV. In Section V, we present some constructions to obtain several infinite families of optimal $t$-TS, and also establish a constructive lower bound for general $t$-TS. Finally, we conclude this paper in Section VI.

\section{Preliminaries}
In this paper, we use the definitions of TS, IPPS and CFF from the viewpoint of set systems. A \textit{set system} is a pair $(\mathcal{X},\mathcal{B})$ where $\mathcal{X}$ is a set of elements called \textit{points} and $\mathcal{B}$ is a collection of subsets of $\mathcal{X}$ called \textit{blocks}. We focus on the case that every block has the same size, that is, the \textit{uniform} set systems. Denote $\binom{\mathcal{X}}{k}$ as the collection of all $k$-subsets of $\mathcal{X}$. The definitions of TS, IPPS and CFF are stated as follows.

\begin{definition}\label{def}
Suppose $(\mathcal{X},\mathcal{B})$ is a set system with $\mathcal{B}\subseteq \binom{\mathcal{X}}{w}$, $|\mathcal{X}|=v$, and $|\mathcal{B}|=M$. Then
\begin{description}
  \item[(1)] $(\mathcal{X},\mathcal{B})$ is a \textit{$t$-traceability scheme} $t$-TS$(w, M, v)$ provided that for every choice of $s\le t$ blocks $B_1,B_2,\ldots,B_s\in \mathcal{B}$ and for any $w$-subset $T\subseteq \bigcup_{1\le j\le s}B_j$, there does not exist a block $B\in \mathcal{B}\setminus\{B_1,B_2,\ldots,B_s\}$ such that $|T\cap B|\ge |T\cap B_j|$ for all $1\le j\le s$.
  \item[(2)] $(\mathcal{X},\mathcal{B})$ is a \textit{$t$-parent-identifying set system} $t$-IPPS$(w,M,v)$ provided that for any $w$-subset $T\subseteq \mathcal{X}$, either $P_t(T)$ is empty, or
             \begin{equation*}
             \bigcap_{\mathcal{P}\in P_t(T)}\mathcal{P}\ne \emptyset,
             \end{equation*}
              where
             \begin{equation*}
              P_t(T)=\{\mathcal{P}\subseteq \mathcal{B}:\, |\mathcal{P}|\le t,\ T\subseteq \bigcup_{B\in \mathcal{P}}B\}.
             \end{equation*}
  \item[(3)] $(\mathcal{X},\mathcal{B})$ is a \textit{$t$-cover-free family} $t$-CFF$(w, M, v)$ provided that for any $t+1$ distinct blocks $B_0,B_1,\ldots,B_t\in \mathcal{B}$, we have
            \begin{equation*}
             B_0\nsubseteq\bigcup_{1\le i\le t}B_i.
             \end{equation*}
\end{description}
\end{definition}

The parameter $M$ is called the \textit{size} of the set system. We also use $t$-TS$(w,v)$ ($t$-IPPS$(w,v)$, $t$-CFF$(w,v)$, resp.) to replace $t$-TS$(w,M,v)$ ($t$-IPPS$(w,M,v)$, $t$-CFF$(w,M,v)$, resp.) when $M$ is unclear or not necessarily claimed. Denote $M_t(w,v)$ ($N_t(w,v)$, $f_t(w,v)$, resp.) as the maximum size of a $t$-TS$(w,v)$ ($t$-IPPS$(w,v)$, $t$-CFF$(w,v)$, resp.). A $t$-TS$(w,v)$ ($t$-IPPS$(w,v)$, $t$-CFF$(w,v)$, resp.) is called \textit{optimal} if it has size $M_t(w,v)$ ($N_t(w,v)$, $f_t(w,v)$, resp.). Given parameters $t,w$ and $v$, the goal is to explore the exact value of $M_t(w,v)$ ($N_t(w,v)$, $f_t(w,v)$, resp.) and to construct the optimal $t$-TS$(w,v)$ ($t$-IPPS$(w,v)$, $t$-CFF$(w,v)$, resp.).

Bounds of $f_t(w,v)$ have been studied by numerous researchers, see \cite{Erdos1982,Erdos1985,Furedi1996,Zhu2000}, for example. An important notion, which is extremely useful in the process of deriving bounds for CFF in \cite{Erdos1985}, is the own-subset. In a set system $(\mathcal{X},\mathcal{B})$, $B\in \mathcal{B}$, a subset $B_0\subseteq B$ is called a \textit{$|B_0|$-own-subset} of $B$ if for any $B'\in \mathcal{B}\setminus \{B\}$, we have $B_0\nsubseteq B'$.

In \cite{Stinson1998}, Stinson and Wei studied the combinatorial properties of traceability schemes and proved the following lemma.
\begin{lemma}[\cite{Stinson1998}]\label{SWr}
A $t$-TS$(w,v)$ is a $t$-CFF$(w,v)$.
\end{lemma}

Using the above relationship, an upper bound for $t$-TS$(w,v)$ was also given in \cite{Stinson1998} by using an upper bound of CFF in \cite{Erdos1985}.

In fact, Lemma \ref{SWr} can be segmented as follows.

\begin{lemma}[\cite{Collins2009}]\label{CollinsR}
A $t$-TS$(w,v)$ is a $t$-IPPS$(w,v)$, and a $t$-IPPS$(w,v)$ is a $t$-CFF$(w,v)$.
\end{lemma}
Thus we have the following corollary.
\begin{coro}
Let $v\ge w\ge t\ge 2$ be integers. Then $M_t(w,v)\le N_t(w,v)\le f_t(w,v)$.
\end{coro}

Noting that the IPPS defined in Definition \ref{def}(2) requires that $|T|=w$. Modifying it to all the case $|T|\ge w$, we have the following definition of $t$-IPPS$^{*}$.

\begin{definition}
A \textit{$t$-parent-identifying$^{*}$ set system} $t$-IPPS$^{*}$$(w,v)$ is a set system $(\mathcal{X},\mathcal{B})$ such that $\mathcal{B}\subseteq \binom{\mathcal{X}}{w}$ and $|\mathcal{X}|=v$, with the property that for any $T\subseteq \mathcal{X}$ such that $|T|\ge w$, either $P_t(T)$ is empty, or
\begin{equation*}
\bigcap_{\mathcal{P}\in P_t(T)}\mathcal{P}\ne \emptyset,
\end{equation*}
where
\begin{equation*}
P_t(T)=\{\mathcal{P}\subseteq \mathcal{B}:\, |\mathcal{P}|\le t,\ T\subseteq \bigcup_{B\in \mathcal{P}}B\}.
\end{equation*}
\end{definition}

Considering the relationship between IPPS and IPPS$^{*}$, we have the following lemma.
\begin{lemma}\label{IPPSrSTAR}
A set system $(\mathcal{X},\mathcal{B})$ is a $t$-IPPS$(w,v)$ if and only if it is a $t$-IPPS$^{*}$$(w,v)$.
\end{lemma}
\pf The sufficiency directly follows from their definitions. We focus on the necessity.
Suppose $(\mathcal{X},\mathcal{B})$ is a $t$-IPPS$(w,v)$, we would like to show that it is also a $t$-IPPS$^{*}$$(w,v)$. Consider any $T\subseteq \mathcal{X}$ with $|T|\ge w$ and $P_t(T)\ne \emptyset$. Choosing a $w$-subset $T_0\subseteq T$, we have
\begin{equation*}
P_t(T)\subseteq P_t(T_0),
\end{equation*}
since for any $\mathcal{P}\in P_t(T)$, we have $T\subseteq \bigcup_{B\in \mathcal{P}}B$ and then $T_0\subseteq T\subseteq \bigcup_{B\in \mathcal{P}}B$, which implies that $\mathcal{P}\in P_t(T_0)$. By the definition of $t$-IPPS$(w,v)$, we have
\begin{equation*}
\bigcap_{\mathcal{P}\in P_t(T_0)}\mathcal{P}\ne \emptyset.
\end{equation*}
Hence we have
\begin{equation*}
\emptyset\ne \bigcap_{\mathcal{P}\in P_t(T_0)}\mathcal{P}\subseteq \bigcap_{\mathcal{P}\in P_t(T)}\mathcal{P}.
\end{equation*}
Thus $(\mathcal{X},\mathcal{B})$ is a $t$-IPPS$^{*}$$(w,v)$, and the necessity follows.\qed

\section{Upper bounds for $t$-IPPS}
\subsection{A known upper bound for $t$-IPPS}
By investigating $\lceil\frac{w}{\lfloor t^2/4\rfloor+\lceil t/2\rceil}\rceil$-own-subsets, Collins gave an upper bound for $t$-IPPS$(w,v)$ as follows.
\begin{theorem}[\cite{Collins2009}]
Let $v\ge w\ge t\ge 2$ be integers. Then
\begin{equation*}
N_t(w,v)\le \frac{\binom{v}{\lceil\frac{w}{\lfloor t^2/4\rfloor+\lceil t/2\rceil}\rceil}}{\binom{\lceil\frac{w}{\lfloor t/2\rfloor+1}\rceil-1}{\lceil\frac{w}{\lfloor t^2/4\rfloor+\lceil t/2\rceil}\rceil-1}}=O(v^{\lceil\frac{w}{\lfloor t^2/4\rfloor+\lceil t/2\rceil}\rceil}).
\end{equation*}
\end{theorem}

\subsection{A new upper bound for $t$-IPPS}
In this subsection, we provide a better upper bound as follows, by showing that blocks of a $t$-IPPS must contain smaller own-subsets.

\begin{theorem}\label{newboundIPPS}
Let $v\ge w\ge t\ge 2$ be integers. Then
\begin{equation*}
N_t(w,v)\le \binom{v}{\lceil\frac{w}{\lfloor t^2/4\rfloor+t}\rceil}=O(v^{\lceil\frac{w}{\lfloor t^2/4\rfloor+t}\rceil}).
\end{equation*}
\end{theorem}

Before proving this theorem, we need the following lemma.
\begin{lemma}\label{newlemma}
Let $(\mathcal{X},\mathcal{B})$ be a $t$-IPPS$(w,v)$. There exists one block $B\in \mathcal{B}$ containing at least one $\lceil\frac{w}{\lfloor t^2/4\rfloor+t}\rceil$-own-subset.
\end{lemma}

\pf Suppose on the contrary that each block $B\in \mathcal{B}$ does not contain any $\lceil\frac{w}{\lfloor t^2/4\rfloor+t}\rceil$-own-subset. That is, for each block $B\in \mathcal{B}$ and each $\lceil\frac{w}{\lfloor t^2/4\rfloor+t}\rceil$-subset $B_0\subseteq B$, there exists another block $B'\in \mathcal{B}\setminus \{B\}$ such that $B_0\subseteq B'$. Then we would like to derive a contradiction with the assumption that $(\mathcal{X},\mathcal{B})$ is a $t$-IPPS$(w,v)$.

Firstly, arbitrarily choose a block $B_1\in \mathcal{B}$, and take a subset $A_1\subseteq B_1$ such that $|A_1|=\lceil\frac{w}{\lfloor t^2/4\rfloor+t}\rceil\lceil\frac{t}{2}\rceil$. By the assumption, $A_1$ can be covered by at most $\lceil\frac{t}{2}\rceil$ distinct blocks in $\mathcal{B}$ other than $B_1$. Denote $\mathcal{C}^{(1)}\subseteq \mathcal{B}\setminus \{B_1\}$ such that $|\mathcal{C}^{(1)}|\le \lceil\frac{t}{2}\rceil$ and $A_1\subseteq \bigcup_{B\in \mathcal{C}^{(1)}}B$. Then take another subset $D_1\subseteq B_1\setminus A_1$ such that $|D_1|=\lceil\frac{w}{\lfloor t^2/4\rfloor+t}\rceil$. With the assumption, there exists another block $B_2\in \mathcal{B}\setminus \{B_1\}$ such that $D_1\subseteq B_2$. Let $i=2$.

While $2\le i\le \lfloor \frac{t}{2}\rfloor$, take a subset $A_i\subseteq B_i\setminus (\cup_{1\le j\le i-1}(A_j\cup D_j))$ such that $|A_i|=\lceil\frac{w}{\lfloor t^2/4\rfloor+t}\rceil\lceil\frac{t}{2}\rceil$. Note that this is doable since the size of $B_i\in \mathcal{B}$ is $w$. With the assumption, each $A_i$ can be covered by at most $\lceil\frac{t}{2}\rceil$ distinct blocks in $\mathcal{B}$ other than $B_i$. Denote $\mathcal{C}^{(i)}\subseteq \mathcal{B}\setminus \{B_i\}$ such that $|\mathcal{C}^{(i)}|\le \lceil\frac{t}{2}\rceil$ and $A_i\subseteq \bigcup_{B\in \mathcal{C}^{(i)}}B$. Note that, some $B_j$, $j\ne i$, may appear in $\mathcal{C}^{(i)}$. This is allowed since it does not increase the number of blocks that we are looking for. After this, we have
\begin{equation*}
B_i\setminus (\bigcup_{1\le j\le i-1}(A_j\cup D_j)\cup A_i)\nsubseteq (\bigcup_{1\le j\le i-1}B_j)\setminus (\bigcup_{1\le j\le i-1}(A_j\cup D_j)\cup A_i),
\end{equation*}
since if not, $B_i$ would be covered by at most $t$ other blocks $\{B_j: 1\le j\le i-1\}\cup \mathcal{C}^{(i)}$, which contradicts to Lemma \ref{CollinsR}. This allows us to take another subset $D_i\subseteq B_i\setminus (\bigcup_{1\le j\le i-1}(A_j\cup D_j)\cup A_i)$ such that $|D_i|=\lceil\frac{w}{\lfloor t^2/4\rfloor+t}\rceil$ and
\begin{equation*}
D_i\nsubseteq (\bigcup_{1\le j\le i-1}B_j)\setminus (\bigcup_{1\le j\le i-1}(A_j\cup D_j)\cup A_i).
\end{equation*}
With the assumption, there exists another block $B_{i+1}\in \mathcal{B}\setminus \{B_j: 1\le j\le i\}$ such that $D_i\subseteq B_{i+1}$. Let $i=i+1$ and continually execute the while loop of this paragraph.

The above while loop stops at $i=\lfloor \frac{t}{2}\rfloor+1$, then take $A_{\lfloor \frac{t}{2}\rfloor+1}\subseteq B_{\lfloor \frac{t}{2}\rfloor+1}\setminus (\cup_{1\le j\le \lfloor \frac{t}{2}\rfloor}(A_j\cup D_j))$ such that
\begin{equation*}
|A_{\lfloor \frac{t}{2}\rfloor+1}|=w-\lceil\frac{w}{\lfloor t^2/4\rfloor+t}\rceil\lceil\frac{t}{2}\rceil \lfloor \frac{t}{2}\rfloor-\lceil\frac{w}{\lfloor t^2/4\rfloor+t}\rceil \lfloor \frac{t}{2}\rfloor.
\end{equation*}
Clearly, $|A_{\lfloor \frac{t}{2}\rfloor+1}|\le \lceil\frac{w}{\lfloor t^2/4\rfloor+t}\rceil\lceil\frac{t}{2}\rceil$. 
With the assumption, $A_{\lfloor \frac{t}{2}\rfloor+1}$ can be covered by at most $\lceil\frac{t}{2}\rceil$ distinct blocks in $\mathcal{B}$ other than $B_{\lfloor \frac{t}{2}\rfloor+1}$. Denote $\mathcal{C}^{({\lfloor \frac{t}{2}\rfloor+1})}\subseteq \mathcal{B}\setminus \{B_{\lfloor \frac{t}{2}\rfloor+1}\}$ such that $|\mathcal{C}^{({\lfloor \frac{t}{2}\rfloor+1})}|\le \lceil\frac{t}{2}\rceil$ and $A_{\lfloor \frac{t}{2}\rfloor+1}\subseteq \bigcup_{B\in \mathcal{C}^{({\lfloor \frac{t}{2}\rfloor+1})}}B$.

Now, we have already taken ${\lfloor \frac{t}{2}\rfloor+1}$ distinct blocks $B_1,\ldots,B_{\lfloor \frac{t}{2}\rfloor+1}\in \mathcal{B}$. Denote
\begin{equation*}
T=\bigcup_{1\le j\le {\lfloor \frac{t}{2}\rfloor}} (A_j\cup D_j)\cup A_{\lfloor \frac{t}{2}\rfloor+1}.
\end{equation*}
Clearly, $T\subseteq \mathcal{X}$ and $|T|=w$. Moreover,
\begin{equation*}
T\subseteq \bigcup_{1\le j\le {\lfloor \frac{t}{2}\rfloor+1}}B_j,
\end{equation*}
that is, $\mathcal{P}_0:=\{B_1,\ldots,B_{\lfloor \frac{t}{2}\rfloor+1}\}\in P_t(T)$.

On the other hand, for each $1\le i\le {\lfloor \frac{t}{2}\rfloor+1}$, we have
\begin{equation*}
T\subseteq (\bigcup_{B\in \mathcal{C}^{(i)}}B)\cup (\bigcup_{1\le j\le {\lfloor \frac{t}{2}\rfloor+1},\atop j\ne i}B_{j}),
\end{equation*}
that is, each $\mathcal{P}_i:=\mathcal{C}^{(i)}\cup \{B_j:\, 1\le j\le {\lfloor \frac{t}{2}\rfloor+1},\, j\ne i\}\subseteq \mathcal{B}$ covers $T$. Moreover, $|\mathcal{P}_i|\le \lceil\frac{t}{2}\rceil+\lfloor \frac{t}{2}\rfloor=t$. That is,
\begin{equation*}
\{\mathcal{P}_0,\mathcal{P}_1,\ldots,\mathcal{P}_{\lfloor \frac{t}{2}\rfloor+1}\}\subseteq P_t(T).
\end{equation*}
However,
\begin{equation*}
\bigcap_{0\le i\le {\lfloor \frac{t}{2}\rfloor+1}}\mathcal{P}_i= \emptyset,
\end{equation*}
which implies
\begin{equation*}
\bigcap_{\mathcal{P}\in P_t(T)}\mathcal{P}= \emptyset.
\end{equation*}
Hence $(\mathcal{X},\mathcal{B})$ is not a $t$-IPPS$(w,v)$, a contradiction to the assumption. Therefore the lemma follows.\qed

\textbf{Proof of Theorem \ref{newboundIPPS}}. Suppose $(\mathcal{X},\mathcal{B})$ is a $t$-IPPS$(w,v)$. By Lemma \ref{newlemma}, there exists one block $B\in \mathcal{B}$ which contains at least one $\lceil\frac{w}{\lfloor t^2/4\rfloor+t}\rceil$-own-subset. Delete this $B$ from $\mathcal{B}$. The resulting $\mathcal{B}\setminus \{B\}$ is still a $t$-IPPS$(w,v)$. Applying Lemma \ref{newlemma} repeatedly, we can successively delete blocks, which contain at least one $\lceil\frac{w}{\lfloor t^2/4\rfloor+t}\rceil$-own-subset, from the newly obtained $t$-IPPS$(w,v)$. Note that there are $\binom{v}{\lceil\frac{w}{\lfloor t^2/4\rfloor+t}\rceil}$ distinct $\lceil\frac{w}{\lfloor t^2/4\rfloor+t}\rceil$-subsets from $\mathcal{X}$, and each $\lceil\frac{w}{\lfloor t^2/4\rfloor+t}\rceil$-subset, as own-subset of some block, can be deleted at most once. Hence
\begin{equation*}
|\mathcal{B}|\le \binom{v}{\lceil\frac{w}{\lfloor t^2/4\rfloor+t}\rceil},
\end{equation*}
and the theorem follows.\qed

\subsection{Analysis of the new upper bound for $t$-IPPS}
In \cite{Barg2001}, Barg, Cohen, Encheva, Kabatiansky and Z\'{e}mor exploited the notion of \textit{minimal forbidden configuration} to study IPP codes. Here we use similar notions to analyze IPPS. In a set system $(\mathcal{X},\mathcal{B})$, let $\mathcal{F}=\{\mathcal{F}_1,\ldots,\mathcal{F}_m\}$ be a collection of subsets of blocks with $\mathcal{F}_i\subseteq \mathcal{B},\ |\mathcal{F}_i|\le t$, $i=1,\ldots,m$. Then $\mathcal{F}$ is called a \textit{configuration} if it has an empty intersection, $\cap_{1\le i\le m}\mathcal{F}_i=\emptyset$. $\mathcal{F}$ is called a \textit{minimal configuration} if it is minimal under inclusion, that is, $\cap_{1\le j\le m,\atop {j\ne i}}\mathcal{F}_j\ne \emptyset$ for any $i=1,\ldots,m$. In \cite{Barg2001}, Barg et al. proved the following lemma for the size of a minimal configuration, which was also showed by Staddon, Stinson and Wei in \cite{Staddon2001}.

\begin{lemma}[\cite{Barg2001}, \cite{Staddon2001}]\label{minmalconf}
Let $\mathcal{F}$ be a minimal configuration. Then $|\cup_{\mathcal{F}_i\in \mathcal{F}}\mathcal{F}_i|\le \lfloor(\frac{t}{2}+1)^2\rfloor$.
\end{lemma}

Recall that in the proof of Lemma \ref{newlemma}, we tried to construct a suitable $T$ for which $\mathcal{P}_t(T)$ is a configuration. Note that the configuration $\{\mathcal{P}_0,\mathcal{P}_1,\ldots,\mathcal{P}_{\lfloor \frac{t}{2}\rfloor+1}\}$ satisfies that
\begin{equation*}
|\bigcup_{0\le i\le \lfloor \frac{t}{2}\rfloor+1}\mathcal{P}_i|\le (\lfloor \frac{t}{2}\rfloor+1)+\lceil\frac{t}{2}\rceil(\lfloor \frac{t}{2}\rfloor+1)=\lfloor(\frac{t}{2}+1)^2\rfloor,
\end{equation*}
which matches the bound of size of a minimal configuration in Lemma \ref{minmalconf}. We conjecture that, by using the method of constructing $T$ in Lemma \ref{newlemma}, the corresponding $\{\mathcal{P}_0,\mathcal{P}_1,\ldots,\mathcal{P}_{\lfloor \frac{t}{2}\rfloor+1}\}$ is a (minimal) configuration with the smallest size, and consequently the new upper bound $\binom{v}{\lceil\frac{w}{\lfloor t^2/4\rfloor+t}\rceil}$ is the exact upper bound for $t$-IPPS$(w,v)$, up to a constant depending only on $w$ and $t$.

\section{Upper bounds for $t$-TS}

From this section, we focus on TS, a special kind of IPPS with a more efficient tracing algorithm.

\subsection{Known upper bounds for $t$-TS}
In \cite{Stinson1998}, Stinson and Wei provided an upper bound for $t$-TS as follows.
\begin{theorem}[\cite{Stinson1998}]\label{SWb}
For any $v\ge w\ge t\ge 2$, we have
\begin{equation*}
M_t(w,v)\le \binom{v}{\lceil w/{t}\rceil}/\binom{w-1}{\lceil w/t\rceil-1}.
\end{equation*}
\end{theorem}

By investigating $\lceil w/{t^2}\rceil$-own-subsets, Collins \cite{Collins2009} improved the above upper bound for $t$-TS$(w,v)$ as follows.
\begin{theorem}[\cite{Collins2009}]\label{Collinsb}
For any $v\ge w\ge t\ge 2$, we have
\begin{equation*}
M_t(w,v)\le \binom{v}{\lceil w/{t^2}\rceil}.
\end{equation*}
\end{theorem}

However, in the literature, no construction can produce $t$-TS$(w,v)$ achieving the above upper bound. As a matter of fact, this bound is still not tight. In the next subsection, we show an interesting discovery of the relationship between TS and CFF. Based on this important discovery, new upper bounds for $t$-TS$(w,v)$ are derived, which are great improvements on the previously known upper bounds. Moreover, we describe several constructions in Section V, which can produce infinite families of $t$-TS$(w,v)$ achieving our new upper bounds.

\subsection{A new relationship between TS and CFF}

To show our new upper bounds, we firstly state the following interesting discovery. As far as we know, this is the first relationship between two kinds of anti-collusion schemes which strengthens the strength from $t$ to $t^2$.
\begin{lemma}\label{newr}
A $t$-TS$(w,v)$ is a $t^2$-CFF$(w,v)$.
\end{lemma}
\pf Assume that $(\mathcal{X},\mathcal{B})$ is a $t$-TS$(w,v)$. We would like to show that $(\mathcal{X},\mathcal{B})$ is also a $t^2$-CFF$(w,v)$.
Suppose, on the contrary, that $(\mathcal{X},\mathcal{B})$ is not a $t^2$-CFF$(w,v)$. By the definition of $t^2$-CFF, there exists $B_0\in \mathcal{B}$ which can be covered by the union of some other $t^2$ blocks from $\mathcal{B}\setminus \{B_0\}$. Denote $\mathcal{B}_0$ as the collection of such $t^2$ blocks.

By the pigeonhole principle, there exists at least one block $B\in \mathcal{B}_0$ such that $|B_0\cap B|\ge \lceil\frac{w}{t^2}\rceil$. Denote
\begin{equation*}
\max\{|B_0\cap B|:\, B\in \mathcal{B}_0\}=\lceil\frac{w}{t^2}\rceil +\sigma_1,
\end{equation*}
where $0\le \sigma_1<w-\lceil\frac{w}{t^2}\rceil$. Assume that $B_1\in \mathcal{B}_0$ satisfies $|B_0\cap B_1|=\lceil\frac{w}{t^2}\rceil +\sigma_1$.

For $2\le i\le t$, choose $B_i\in \mathcal{B}_0\setminus \{B_j: 1\le j\le i-1\}$ such that
\begin{equation*}
|B_i\cap (B_0\setminus \cup_{1\le j\le i-1}B_j)|=\sigma_i>0,
\end{equation*}
and
\begin{equation*}
\sum_{2\le i\le t}\sigma_i\ge \frac{1}{t+1}(w-\lceil\frac{w}{t^2}\rceil-\sigma_1). \tag{$1$}
\end{equation*}

The above $t$ blocks, satisfying $\sigma_i>0$ for $2\le i\le t$, are available, since if not, then $B_0$ would be covered by at most other $t$ distinct blocks, which contradicts Lemma \ref{SWr}. Inequality $(1)$ can be realized by using the pigeonhole principle to each block $B_i$, $2\le i\le t$. The reader is referred to Appendix for the detailed computation.

By the assumption, for $2\le i\le t$, we have
\begin{equation*}
\sigma_i\le |B_i\cap B_0|\le |B_1\cap B_0|=\lceil\frac{w}{t^2}\rceil+\sigma_1.
\end{equation*}

Note that the intersection of $B_j$ and $B_k$, $1\le j<k\le t$, may be not contained in $B_0$. If
\begin{equation*}
|(\bigcup_{1\le i<j\le t}B_i\cap B_j) \setminus B_0|\ge w-(\lceil\frac{w}{t^2}\rceil+\sum_{1\le i\le t}\sigma_i),
\end{equation*}
then any $t$ of $B_0,B_1,\ldots,B_t$ can generate the same $w$-subset of
\begin{equation*}
\bigcup_{0\le i<j\le t}(B_i\cap B_j),
\end{equation*}
where $|\bigcup_{0\le i<j\le t}(B_i\cap B_j)|\ge w$, which contradicts Lemma \ref{CollinsR}. So
\begin{equation*}
0\le|(\bigcup_{1\le i<j\le t}B_i\cap B_j) \setminus B_0|< w-(\lceil\frac{w}{t^2}\rceil+\sum_{1\le i\le t}\sigma_i).
\end{equation*}
For each $1\le i\le t$, we have
\begin{equation*}
\begin{split}
|B_i\setminus \bigcup_{0\le j\le t,\atop j\ne i}B_j|&\ge w-|B_i\cap B_0|-|(\bigcup_{1\le j\le t,\atop j\ne i}B_i\cap B_j)\setminus B_0|\\
&>w-(\lceil\frac{w}{t^2}\rceil+\sigma_1)-[w-(\lceil\frac{w}{t^2}\rceil+\sum_{1\le i\le t}\sigma_i)]\\
&=\sum_{2\le i\le t}\sigma_i.
\end{split}
\end{equation*}
That is, each block $B_i$, $1\le i\le t$, has more than $\sum_{2\le i\le t}\sigma_i$ points which are not contained in any other $B_j$ for $0\le j\le t$ and $j\ne i$.
Now choose a set $A\subseteq \bigcup_{1\le i\le t}B_i$ as follows: in each $B_i$, $1\le i\le t$, take all the points in $B_0\cap B_i$ and $\sum_{2\le i\le t}\sigma_i$ points from $B_i\setminus \bigcup_{0\le j\le t,\atop j\ne i}B_j$. Then the size of $A$ is
\begin{equation*}
\begin{split}
|A|&=(\lceil\frac{w}{t^2}\rceil +\sum_{1\le i\le t}\sigma_i)+t\sum_{2\le i\le t}\sigma_i\\
&=\lceil\frac{w}{t^2}\rceil +\sigma_1+(t+1)\sum_{2\le i\le t}\sigma_i\\
&\ge\lceil\frac{w}{t^2}\rceil +\sigma_1+(t+1)\frac{1}{t+1}(w-\lceil\frac{w}{t^2}\rceil-\sigma_1)\\
&=w,
\end{split}
\end{equation*}
where the inequality follows from $(1)$.

Thus $B_1,\ldots,B_t$ can generate a descendant $F$ by taking a $w$-subset $F\subseteq A$ which satisfies $B_0\cap B_i\subseteq F$ for all $1\le i\le t$. Hence we have
\begin{equation*}
|B_i\cap F|\le |B_0\cap B_i|+\sum_{2\le i\le t}\sigma_i\le \lceil\frac{w}{t^2}\rceil+\sum_{1\le i\le t}\sigma_i,\ 1\le i\le t,
\end{equation*}
and
\begin{equation*}
|B_0\cap F|= \lceil\frac{w}{t^2}\rceil +\sum_{1\le i\le t}\sigma_i,\\
\end{equation*}
which contradicts the definition of $t$-TS$(w,v)$.

Therefore, for any $B\in \mathcal{B}$, it can not be covered by the union of any other $t^2$ blocks from $\mathcal{B}\setminus \{B\}$. The lemma follows. \qed

In \cite{Erdos1985}, Erd\"{o}s, Frankl and F\"{u}redi proved the following result.
\begin{lemma}[\cite{Erdos1985}]\label{EFFno}
Let $(\mathcal{X},\mathcal{B})$ be a $t$-CFF$(w,v)$, and $B\in \mathcal{B}$. The number of $\lceil {w}/{t}\rceil$-own-subsets in $B$ is at least $\binom{w-1}{\lceil w/{t}\rceil-1}$.
\end{lemma}

Combining Lemma \ref{newr} and Lemma \ref{EFFno}, we have the following lemma.
\begin{lemma}\label{newno}
Let $(\mathcal{X},\mathcal{B})$ be a $t$-TS$(w,v)$, and $B\in \mathcal{B}$. The number of $\lceil {w}/{t^2}\rceil$-own-subsets in $B$ is at least $\binom{w-1}{\lceil w/{t^2}\rceil-1}$.
\end{lemma}

\subsection{A new upper bound for general $t$-TS}

In this part, we prove the following new upper bound for $t$-TS.
\begin{theorem}\label{newgb}
For any $v\ge w\ge t\ge 2$, we have
\begin{equation*}
M_t(w,v)\le \frac{\binom{v}{\lceil w/{t^2}\rceil}-\binom{w-1}{\lceil w/{t^2}\rceil}}{\binom{w-1}{\lceil w/{t^2}\rceil-1}}.
\end{equation*}
\end{theorem}

\pf Suppose $(\mathcal{X},\mathcal{B})$ is a $t$-TS$(w,M,v)$. Denote $\binom{\mathcal{X}}{\lceil {w}/{t^2}\rceil}$ as the collection of all $\lceil {w}/{t^2}\rceil$-subsets of $\mathcal{X}$. Clearly, $|\binom{\mathcal{X}}{\lceil {w}/{t^2}\rceil}|=\binom{v}{\lceil {w}/{t^2}\rceil}$. The following proof is to double count the set $\{(T,B):\, T\in \binom{\mathcal{X}}{\lceil {w}/{t^2}\rceil},\, B\in \mathcal{B},\, T\subseteq B\}$. Denote
\begin{equation*}
\Sigma:=|\{(T,B):\, T\in \binom{\mathcal{X}}{\lceil {w}/{t^2}\rceil},\, B\in \mathcal{B},\, T\subseteq B\}|.
\end{equation*}
Then we have
\begin{equation*}
\begin{split}
\Sigma&=\sum\limits_{T\in \binom{\mathcal{X}}{\lceil {w}/{t^2}\rceil}}\ \sum\limits_{B\in \mathcal{B}\atop \text{s.t. } T\subseteq B}1\ =\sum\limits_{B\in \mathcal{B}}\ \sum\limits_{T\in \binom{\mathcal{X}}{\lceil {w}/{t^2}\rceil}\atop \text{s.t. } T\subseteq B}1.
\end{split}\tag{$2$}
\end{equation*}
On one hand, fixing $B\in \mathcal{B}$, we have $\sum\limits_{T\in \binom{\mathcal{X}}{\lceil {w}/{t^2}\rceil}\atop \text{s.t. } T\subseteq B}1=\binom{w}{\lceil {w}/{t^2}\rceil}$. Then
\begin{equation*}
\Sigma=\sum\limits_{B\in \mathcal{B}}\binom{w}{\lceil {w}/{t^2}\rceil}=M\binom{w}{\lceil {w}/{t^2}\rceil}.\tag{$3$}
\end{equation*}
On the other hand, fixing $T\in \binom{\mathcal{X}}{\lceil {w}/{t^2}\rceil}$, there are the following two possible cases:
\begin{description}
  \item(a) if $T$ is a $\lceil {w}/{t^2}\rceil$-own-subset of some block $B\in \mathcal{B}$, then we have $\sum\limits_{B\in \mathcal{B}\atop \text{s.t. } T\subseteq B}1=1$;
  \item(b) if $T$ is not a $\lceil {w}/{t^2}\rceil$-own-subset of any block $B\in \mathcal{B}$, then we have $\sum\limits_{B\in \mathcal{B}\atop \text{s.t. } T\subseteq B}1\le M$.
\end{description}

For any $B\in \mathcal{B}$, denote $\mathcal{O}(B)$ as the collection of all $\lceil {w}/{t^2}\rceil$-own-subsets of the block $B$.
By Lemma \ref{newno}, we have $|\mathcal{O}(B)|\ge \binom{w-1}{\lceil w/{t^2}\rceil-1}$. Without loss of generality, we assume that
\begin{equation*}
|\bigcup_{B\in \mathcal{B}}\mathcal{O}(B)|=M\binom{w-1}{\lceil w/{t^2}\rceil-1}+\sigma,\ \sigma\ge 0.
\end{equation*}
Then the number of $T\in \binom{\mathcal{X}}{\lceil {w}/{t^2}\rceil}$, satisfying the condition of case (b), is at most
\begin{equation*}
\binom{v}{\lceil w/{t^2}\rceil}-M\binom{w-1}{\lceil w/{t^2}\rceil-1}-\sigma.
\end{equation*}
Thus by the first equality of $(2)$,
\begin{equation*}
\begin{split}
\Sigma&\le [M\binom{w-1}{\lceil w/{t^2}\rceil-1}+\sigma]+M[\binom{v}{\lceil w/{t^2}\rceil}-M\binom{w-1}{\lceil w/{t^2}\rceil-1}-\sigma]\\
&=M [\binom{v}{\lceil w/{t^2}\rceil}-M\binom{w-1}{\lceil w/{t^2}\rceil-1}+\binom{w-1}{\lceil w/{t^2}\rceil-1}]-(M-1)\sigma\\
&\le M [\binom{v}{\lceil w/{t^2}\rceil}-M\binom{w-1}{\lceil w/{t^2}\rceil-1}+\binom{w-1}{\lceil w/{t^2}\rceil-1}].
\end{split} \tag{$4$}
\end{equation*}
From $(3)$ and $(4)$, we have
\begin{equation*}
M\binom{w}{\lceil w/{t^2}\rceil}\le M[\binom{v}{\lceil w/{t^2}\rceil}-M\binom{w-1}{\lceil w/{t^2}\rceil-1}+\binom{w-1}{\lceil w/{t^2}\rceil-1}],
\end{equation*}
which implies
\begin{equation*}
M\le \frac{\binom{v}{\lceil w/{t^2}\rceil}-\binom{w}{\lceil w/{t^2}\rceil}}{\binom{w-1}{\lceil w/{t^2}\rceil-1}}+1= \frac{\binom{v}{\lceil w/{t^2}\rceil}-\binom{w-1}{\lceil w/{t^2}\rceil}}{\binom{w-1}{\lceil w/{t^2}\rceil-1}},
\end{equation*}
as desired.\qed

\vskip 0.5cm
In \cite{Erdos1985}, Erd\"{o}s, Frankl and F\"{u}redi obtained the following upper bound for $t$-CFF$(w,v)$.
\begin{theorem}[\cite{Erdos1985}]\label{EFFCFFb}
For any $v\ge w\ge t\ge 2$, we have
\begin{equation*}
f_{t}(w,v)\le \frac{\binom{v}{\lceil w/{t}\rceil}}{\binom{w-1}{\lceil w/{t}\rceil-1}}.
\end{equation*}
\end{theorem}

Note that the double-counting technique used in the proof of Theorem \ref{newgb} also can be applied to derive an upper bound for $t$-CFF$(w,v)$. More precisely, we have the following upper bound for $t$-CFF$(w,v)$, which is slightly better than that in Theorem \ref{EFFCFFb}.
\begin{theorem}\label{newCFFb}
For any $v\ge w\ge t\ge 2$, we have
\begin{equation*}
f_{t}(w,v)\le \frac{\binom{v}{\lceil w/{t}\rceil}-\binom{w-1}{\lceil w/{t}\rceil}}{\binom{w-1}{\lceil w/{t}\rceil-1}}.
\end{equation*}
\end{theorem}
\pf The proof is similar to that of Theorem \ref{newgb}, and we omit it here.\qed

\subsection{A better upper bound for $t$-TS in some special cases}

Besides the general upper bound for $t$-TS$(w,v)$ that we proved in the previous subsection, a better upper bound for several special cases can be obtained. Erd\"{o}s, Frankl and F\"{u}redi \cite{Erdos1985} provided the following bound for $f_r(w,v)$.

\begin{theorem}[\cite{Erdos1985}]\label{EFFsb}
 Let $w=r(\lceil\frac{w}{r}\rceil-1)+1+d$ where $0\le d\le r-1$. Then for $v>2d\lceil\frac{w}{r}\rceil\binom{w}{\lceil {w}/{r}\rceil}$,
\begin{equation*}
f_r(w,v)\le \frac{\binom{v-d}{\lceil w/r\rceil}}{\binom{w-d}{\lceil w/r\rceil}}
\end{equation*}
holds in the following cases:
\begin{equation*}
\text{(a)}\ d=0,1,\quad \text{(b)}\ d<r/(2\lceil\frac{w}{r}\rceil^2),\quad \text{(c)}\  \lceil\frac{w}{r}\rceil=2\ \text{and}\ d< \lceil 2r/3\rceil.
\end{equation*}
\end{theorem}
By using the relationship in Lemma \ref{newr}, we have the following bound for $t$-TS.
\begin{theorem}\label{newsb}
Let $w=t^2(\lceil\frac{w}{t^2}\rceil-1)+1+d$ where $0\le d\le t^2-1$. Then for $v>2d\lceil\frac{w}{t^2}\rceil\binom{w}{\lceil {w}/{t^2}\rceil}$,
\begin{equation*}
M_t(w,v)\le \frac{\binom{v-d}{\lceil w/{t^2}\rceil}}{\binom{w-d}{\lceil w/{t^2}\rceil}}
\end{equation*}
holds in the following cases:
\begin{equation*}
\text{(a)}\ d=0,1,\quad \text{(b)}\ d<t^2/(2\lceil\frac{w}{t^2}\rceil^2),\quad \text{(c)}\  \lceil\frac{w}{t^2}\rceil=2\ \text{and}\ d< \lceil 2t^2/3\rceil.
\end{equation*}
\end{theorem}
\pf This theorem follows from Lemma \ref{newr} and Theorem \ref{EFFsb}.\qed

\section{Lower bounds for $t$-TS$(w,v)$}
\subsection{When $w\le t^2$}

\begin{theorem}\label{newssbo}
For any $v\ge w\ge t\ge 2$, we have
\begin{equation*}
M_t(w,v)\ge v-w+1.
\end{equation*}
\end{theorem}
\pf We prove this theorem by providing a construction. Suppose $\mathcal{X}$ is a $v$-set of points. Arbitrarily choose a $(w-1)$-subset $\Delta\subseteq \mathcal{X}$. Define
\begin{equation*}
B_j:=(\{j\}\cup \Delta)\subseteq \mathcal{X},\ \text{for}\ \forall j\in \mathcal{X}\setminus \Delta,
\end{equation*}
and denote $\mathcal{B}:=\{B_j:\, j\in \mathcal{X}\setminus \Delta\}$. Then $(\mathcal{X},\mathcal{B})$ is a $t$-TS$(w,v)$ for any $t\ge 2$, since besides the common subset $\Delta$, each block possesses a unique point. So
\begin{equation*}
M_t(w,v)\ge |\mathcal{B}|=v-(w-1)=v-w+1,
\end{equation*}
as desired. \qed

Considering the case $w\le t^2$ in Theorem \ref{newgb}, we have the following corollary.
\begin{coro}\label{newssb}
For any $v\ge w\ge t\ge 2$ with $w\le t^2$, we have
\begin{equation*}
M_t(w,v)\le v-w+1.
\end{equation*}
\end{coro}

Combining Theorem \ref{newssb} and Corollary \ref{newssb}, we have the following result for the case $w\le t^2$.

\begin{coro}
For any $v\ge w\ge t\ge 2$ with $w\le t^2$, we have
\begin{equation*}
M_t(w,v)= v-w+1.
\end{equation*}
\end{coro}

Note that the size of $t$-TS$(w,v)$ we obtained in Theorem \ref{newssbo} is far from the upper bound in Theorem \ref{newgb} for $w>t^2$. So we explore other constructions in the next subsection.

\subsection{Constructions from combinatorial designs}
Combinatorial structures are often used to construct various configurations in coding theory. In this subsection, we use combinatorial designs to construct $t$-TS$(w,v)$. The definition of $\tau$-design can be stated as follows.

\begin{definition}
A $\tau$-$(v,w,\lambda)$ \textit{design} is a set system $(\mathcal{X},\mathcal{B})$, where $\mathcal{X}$ is a $v$-set of points
and $\mathcal{B}$ is a collection of $w$-subsets of $\mathcal{X}$ (blocks), with the property that every $\tau$-subset
of $\mathcal{X}$ is contained in exactly $\lambda$ blocks. The parameter $\lambda$ is the index of the design.
\end{definition}

We focus on using $\tau$-$(v,w,1)$ design to construct $t$-TS$(w,v)$. Clearly, the number of blocks in a $\tau$-$(v,w,1)$ design is ${\binom{v}{\tau}}/{\binom{w}{\tau}}$.

In \cite{Stinson1998}, Stinson and Wei gave a method of using $\tau$-$(v,w,\lambda)$ design to construct $t$-TS$(w,v)$ as follows.
\begin{theorem}[\cite{Stinson1998}]\label{SWdesign}
If there exists a $\tau$-$(v,w,1)$ design, then there exists a $t$-TS$(w,v)$ with cardinality $\binom{v}{\tau}/\binom{w}{\tau}$, where $t=\lfloor\sqrt{(w-1)/(\tau-1)}\rfloor$.
\end{theorem}

We provide a generalized construction as follows.
\begin{theorem}\label{newdesign}
Let $w\equiv d+1\,({\rm mod}\,t^2)$ where $0\le d\le t^2-1$ and let $\tau=\lceil {w}/{t^2}\rceil$. If there exists a $\tau$-$(v-d,w-d,1)$ design, then there exists a $t$-TS$(w,v)$ with cardinality $\binom{v-d}{\tau}/\binom{w-d}{\tau}$.
\end{theorem}
\pf Let $\mathcal{X}$ be a $v$-set of points. Suppose there exists a $\tau$-$(v-d,w-d,1)$ design $(\mathcal{X}_0,\mathcal{B}_0)$, where $\mathcal{X}_0\subseteq\mathcal{X}$, $|\mathcal{X}\setminus\mathcal{X}_0|=d$. Extending each block $B_0\in\mathcal{B}_0$ to the $d$ points of $\mathcal{X}\setminus\mathcal{X}_0$ to obtain
\begin{equation*}
\mathcal{B}=\{B_0\cup (\mathcal{X}\setminus\mathcal{X}_0):\, B_0\in \mathcal{B}_0\}.
\end{equation*}
Clearly, $|\mathcal{B}|=|\mathcal{B}_0|=\binom{v-d}{\tau}/\binom{w-d}{\tau}$, and $|B|=w$ for any $B\in \mathcal{B}$. We prove that $(\mathcal{X},\mathcal{B})$ is a $t$-TS$(w,v)$.

Suppose $B_1,\ldots,B_s\in \mathcal{B}$, $2\le s\le t$, are $s$ distinct blocks and $B_{s+1}\in \mathcal{B}\setminus \{B_i: 1\le i\le s\}$ is any other block. For any $B\subseteq \cup_{1\le i\le s}B_i$ such that $|B|=w$, we want to show that
\begin{equation*}
|B\cap B_{s+1}|<\max\{|B\cap B_i|:\, 1\le i\le s\}. \tag{$5$}
\end{equation*}
For $1\le i\le s+1$, denote
\begin{equation*}
B'_i=B_i\cap \mathcal{X}_0\in \mathcal{B}_0.
\end{equation*}
Equivalently, $B'_i$ is the corresponding original block in $\mathcal{B}_0$ which is extended to $B_i$.
Denote $B'=B\cap \mathcal{X}_0$. Clearly,
\begin{equation*}
B'\subseteq \bigcup_{1\le i\le s}B'_i. \tag{$6$}
\end{equation*}
Note that $|B'|\ge w-d$ rather than $|B'|=w-d$, so more detailed analyses are required. Assume that
\begin{equation*}
|B\setminus B'|=|B\cap (\mathcal{X}\setminus\mathcal{X}_0)|=\delta,\ 0\le \delta\le d.
\end{equation*}
Then $|B'|=w-\delta$. On the one hand, applying the pigeonhole principle to $(6)$, we have
\begin{equation*}
\max\{|B'\cap B'_i|:\, 1\le i\le s\}\ge \lceil\frac{w-\delta}{s}\rceil\ge \lceil\frac{w-d}{t}\rceil=t(\lceil\frac{w}{t^2}\rceil-1)+1.
\end{equation*}
Since $\mathcal{X}\setminus\mathcal{X}_0$ is contained in each block $B\in \mathcal{B}$, we have $|B\cap B_i|=|B'\cap B'_i|+\delta$. Moreover,
\begin{equation*}
\max\{|B\cap B_i|:\, 1\le i\le s\}\ge t(\lceil\frac{w}{t^2}\rceil-1)+\delta+1.\tag{$7$}
\end{equation*}
On the other hand, since $(\mathcal{X}_0,\mathcal{B}_0)$ is a $\tau$-$(v-d,w-d,1)$ design, we have
\begin{equation*}
\begin{split}
|B'\cap B'_{s+1}|&\le |(\bigcup_{1\le i\le s}B'_i)\cap B'_{s+1}|\\
&=|\bigcup_{1\le i\le s}(B'_i\cap B'_{s+1})|\\
&\le \sum_{1\le i\le s}|B'_i\cap B'_{s+1}|\\
&\le s(\tau-1)\\
&\le t(\lceil\frac{w}{t^2}\rceil-1).
\end{split}
\end{equation*}
Thus
\begin{equation*}
|B\cap B_{s+1}|=|B'\cap B'_{s+1}|+\delta\le t(\lceil\frac{w}{t^2}\rceil-1)+\delta.\tag{$8$}
\end{equation*}
Hence, $(5)$ can be obtained from $(7)$ and $(8)$. With the definition, $(\mathcal{X},\mathcal{B})$ is a $t$-TS$(w,v)$ and the theorem follows.\qed

As can be seen from Theorem \ref{newsb} and Theorem \ref{newdesign}, when all the parameters $v,w,t,d,\tau$ satisfy the conditions therein, the $t$-TS$(w,v)$ constructed from $\tau$-$(v-d,w-d,1)$ design is optimal. From this point of view, the existence of $\tau$-$(v,w,1)$ design is crucial to the existence of optimal $t$-TS. We list several infinite families of optimal $t$-TS$(w,v)$ as follows.

\begin{theorem}\label{design1}
Let $t$ be a prime power and $d$ be an integer such that $0\le d<\lceil 2t^2/3\rceil$. There exists an optimal $t$-TS$(t^2+d+1,\, t^{2n}+t^{2(n-1)}+\cdots+t^2+d+1)$ provided $n\ge 2+\min\{2,d\}$.

\end{theorem}
\pf A $2$-$(q^n+\cdots+q+1,\, q+1,\, 1)$ design exists whenever $q$ is a prime power and $n\ge 2$ \cite{Colbourn2007}. From Theorem \ref{newdesign}, assuming $q=t^2$, for $0\le d\le t^2-1$, there exists a $t$-TS$(t^2+d+1,\, t^{2n}+t^{2(n-1)}+\cdots+t^2+d+1)$. Noting that $\lceil\frac{t^2+d+1}{t^2}\rceil=2$, if $0\le d<\lceil 2t^2/3\rceil$, then this satisfies the condition of case (c) in Theorem \ref{newsb}. It follows that the above $t$-TS$(t^2+d+1,\, t^{2n}+t^{2(n-1)}+\cdots+t^2+d+1)$ is optimal when
\begin{equation*}
t^{2n}+t^{2(n-1)}+\cdots+t^2+d+1>2d(t^2+d)(t^2+d+1),
\end{equation*}
which holds when $n\ge 2+\min\{2,d\}$.\qed

\begin{theorem}\label{design5}
Let $t\ge 2$ be an integer such that $t^2+1$ is a prime power. Let $d$ be an integer such that $0\le d<\lceil 2t^2/3\rceil$. There exists an optimal $t$-TS$(t^2+d+1,\, (t^{2}+1)^n+d)$ provided $n\ge 2+\min\{2,d\}$.
\end{theorem}
\pf A $2$-$(q^n,q,1)$ design exists whenever $q$ is a prime power and $n\ge 2$ \cite{Colbourn2007}. From Theorem \ref{newdesign}, assuming that $q=t^2+1$ is a prime power, for $0\le d\le t^2-1$, there exists a $t$-TS$(t^2+d+1,\, (t^{2}+1)^n+d)$. Noting that $\lceil\frac{t^2+d+1}{t^2}\rceil=2$, if $0\le d<\lceil 2t^2/3\rceil$, then this satisfies the condition of case (c) in Theorem \ref{newsb}. It follows that the above $t$-TS$(t^2+d+1,\, (t^{2}+1)^n+d)$ is optimal when
\begin{equation*}
(t^{2}+1)^n+d>2d(t^2+d)(t^2+d+1),
\end{equation*}
which holds when $n\ge 2+\min\{2,d\}$. \qed

\begin{theorem}\label{design3}
Let $t$ be a positive integer power of $2$, and $d$ be an integer such that $d\in \{0,1\}$ or $0\le d< t^2/18$. There exists an optimal $t$-TS$(2t^2+d+1,\, 2^nt^{2n}+d+1)$ provided $n\ge 2+2\min\{1,d\}$.
\end{theorem}
\pf A $3$-$(q^n+1,q+1,1)$ design exists whenever $q$ is a prime power and $n\ge 2$ \cite{Colbourn2007}. From Theorem \ref{newdesign}, assuming $q=2t^2$ (a power of $2$), for $0\le d\le t^2-1$, there exists a $t$-TS$(2t^2+d+1,\, 2^nt^{2n}+d+1)$. Note that if $d\in \{0,1\}$ or $0\le d< t^2/18$, then this satisfies the condition of case (a) or (b) in Theorem \ref{newsb}. It follows that the above $t$-TS$(2t^2+d+1,\, 2^nt^{2n}+d+1)$ is optimal when
\begin{equation*}
2^nt^{2n}+d+1>d(2t^2+d+1)(2t^2+d)(2t^2+d-1),
\end{equation*}
which holds when $n\ge 2+2\min\{1,d\}$.\qed

\begin{theorem}\label{design4}
Let $t$ be a prime power and $d$ be an integer such that $0\le d\le t^2/4$. There exists an optimal $t$-TS$(t^2+d+1,\, t^6+d+1)$.
\end{theorem}
\pf A $2$-$(q^3+1,q+1,1)$ design exists whenever $q$ is a prime power \cite{Colbourn2007}. From Theorem \ref{newdesign}, assuming $q=t^2$, for $0\le d\le t^2-1$, there exists a $t$-TS$(t^2+d+1,\, t^6+d+1)$. Noting that $\lceil\frac{t^2+d+1}{t^2}\rceil=2$, if $0\le d<\lceil 2t^2/3\rceil$, then this satisfies the condition of case (c) in Theorem \ref{newsb}. It follows that the above $t$-TS$(t^2+d+1,\, t^6+d+1)$ is optimal when
\begin{equation*}
t^6+d+1>2d(t^2+d)(t^2+d+1),
\end{equation*}
which holds when $0\le d\le t^2/4$.\qed

\begin{theorem}\label{design2}
Let $v$ be an integer such that $v\equiv 1,5\,({\rm mod}\,20)$, and $d\in \{0,1,2\}$. There exists an optimal $2$-TS$(5+d,v+d)$ with size $v(v-1)/20$, provided $v>2d^3+18d^2+39d$.
\end{theorem}
\pf A $2$-$(v,5,1)$ design exists whenever $v\equiv 1,5\,({\rm mod}\,20)$ \cite{Colbourn2007}. By Theorem \ref{newdesign}, for $0\le d\le 3$, there exists a $2$-TS$(5+d,v+d)$ with size $v(v-1)/20$. Noting that $\lceil\frac{5+d}{2^2}\rceil=2$, if $0\le d<\lceil 2t^2/3\rceil=3$, then this satisfies the condition of case (c) in Theorem \ref{newsb}. It follows that the above $2$-TS$(5+d,v+d)$ is optimal when $v>2d^3+18d^2+39d$.\qed

In \cite{Stinson1998}, Stinson and Wei gave constructions for $t$-TS via $2$-$(q^2+q+1,q+1,1)$ design and $3$-$(q^2+1,q+1,1)$ design, which are special cases of Theorem \ref{design1} and Theorem \ref{design3}. Particularly, in the above theorems, we show that their constructions and our generalized constructions can produce infinite families of optimal $t$-TS. As is well known, there are plenty of known results on the existence and constructions of $\tau$-design for $\tau=2,3$. However, there are almost no results on the existence of $\tau$-$(v,w,1)$ design when $\tau>5$. From this perspective, it is unreasonable to obtain $t$-TS$(w,v)$ from $\lceil w/t^2\rceil$-design when $\lceil w/t^2\rceil\ge 6$. In the next subsection, we provide a constructive lower bound for general $t$-TS$(w,v)$.

\subsection{A general constructive lower bound}
In \cite{Stinson1998}, Stinson and Wei proposed to use another type of combinatorial designs, packings, to construct TS. The definition of packing is as follows.
\begin{definition}
A $\tau$-$(v,w,\lambda)$ \textit{packing} is a set system $(\mathcal{X},\mathcal{B})$, where $\mathcal{X}$ is a $v$-set of points and $\mathcal{B}$ is a collection of $w$-subsets of $\mathcal{X}$ (blocks), with the property that every $\tau$-subset of $\mathcal{X}$ is contained in at most $\lambda$ blocks.
\end{definition}

\begin{lemma}[\cite{Stinson1998,Blackburn2003}]\label{newlow}
If there exists a $\lceil {w}/{t^2}\rceil$-$(v,w,1)$ packing, then there exists a $t$-TS$(w,v)$.
\end{lemma}

Considering the substance in the above lemma, we have the following constructive lower bound for general $t$-TS.
\begin{theorem}\label{newlb}
For any $v\ge w\ge t\ge 2$, we have
\begin{equation*}
M_t(w,v)\ge {\binom{v}{\lceil w/{t^2}\rceil}}/{\binom{w}{\lceil w/{t^2}\rceil}^2}.
\end{equation*}
\end{theorem}
\pf Firstly, we use the following Algorithm 1.

\begin{algorithm}
\caption{ A construction for $t$-TS$(w,v)$}
\begin{algorithmic}
\INPUT{ $\mathcal{D}_0=\binom{\mathcal{X}}{w}$, $\mathcal{B}=\emptyset$, $i=1$.}
\WHILE{$\mathcal{D}_{i-1}\ne \emptyset$}
\STATE {Choose any block $B_{i}\in \mathcal{D}_{i-1}$;}
\STATE $\mathcal{B}=\mathcal{B}\cup \{B_i\}$;
\STATE $\mathcal{D}_i=\mathcal{D}_{i-1}\setminus\{B\in \mathcal{D}_{i-1}:\, |B\cap B_i|\ge \lceil w/{t^2}\rceil\}$;
\STATE {$i=i+1$.}
\ENDWHILE
\RETURN {$\mathcal{B}$.}
\end{algorithmic}
\end{algorithm}
Clearly, for each $i$, we have
\begin{equation*}
|\mathcal{D}_i\setminus\mathcal{D}_{i-1}|\le\binom{w}{\lceil w/{t^2}\rceil}\binom{v-\lceil w/{t^2}\rceil}{w-\lceil w/{t^2}\rceil}.
\end{equation*}
Hence
\begin{equation*}
\begin{split}
|\mathcal{B}|
&\ge \frac{\binom{v}{w}}{\binom{w}{\lceil w/{t^2}\rceil}\binom{v-\lceil w/{t^2}\rceil}{w-\lceil w/{t^2}\rceil}}
={\binom{v}{\lceil w/{t^2}\rceil}}/{\binom{w}{\lceil w/{t^2}\rceil}^2}.
\end{split}
\end{equation*}
Now, it is sufficient to prove that the collection of blocks $\mathcal{B}$ generated by Algorithm 1 is a $t$-TS$(w,v)$. This follows from Lemma \ref{newlow}, since in Algorithm 1, for any two distinct blocks $B_1,B_2\in \mathcal{B}$, we require that $|B_1\cap B_2|<\lceil w/{t^2}\rceil$, which results in a $\lceil w/{t^2}\rceil$-$(v,w,1)$ packing.

Consequently, we have
\begin{equation*}
M_t(w,v)\ge {\binom{v}{\lceil w/{t^2}\rceil}}/{\binom{w}{\lceil w/{t^2}\rceil}^2},
\end{equation*}
as desired.\qed

We make a remark here that this constructive lower bound has the same order with our general upper bound in Theorem \ref{newgb}.

\section{Conclusions}
In this paper, we explored two structures applied in broadcast encryption systems. For $t$-IPPS, we showed a new upper bound by investigating small own-subsets. For $t$-TS, we firstly found a new relationship between TS and CFF, that is, a $t$-TS is a $t^2$-CFF. Based on this interesting discovery, we derived new upper bounds for $t$-TS. Moreover, we provided several constructions which can produce infinite families of optimal $t$-TS. In other words, our new upper bounds are tight in these cases. Finally, we gave a constructive lower bound for general $t$-TS.

We wonder whether our new upper bounds for $t$-IPPS and $t$-TS are tight for all the other cases, and if it is true, are there deterministic constructions which can (asymptotically) achieve these new upper bounds? These interesting problems are worth investigating in the future.

\appendix
{\bf Proof of inequality $(1)$ in Lemma \ref{newr}.}
By the pigeonhole principle, for each $i$, $2\le i\le t$, there exists at least one block $B_i\in \mathcal{B}_0\setminus \{B_j:\, 1\le j\le i-1\}$ who has no less than $\lceil\frac{|B_0\setminus \cup_{1\le j\le i-1}B_j|}{t^2-i+1}\rceil$ common points with $B_0\setminus \cup_{1\le j\le i-1}B_j$. That is,
\begin{equation*}
|B_i\cap (B_0\setminus \cup_{1\le j\le i-1}B_j)|=\sigma_i\ge \lceil\frac{w-\lceil\frac{w}{t^2}\rceil-\sum_{1\le j\le i-1}\sigma_j}{t^2-i+1}\rceil>0.
\end{equation*}
Then we have
\begin{equation*}
\begin{split}
\sum_{1\le i\le t}\sigma_i&=\sum_{1\le i\le t-1}\sigma_i+\sigma_t\\
&\ge \sum_{1\le i\le t-1}\sigma_i+\frac{1}{t^2-t+1}(w-\lceil\frac{w}{t^2}\rceil-\sum_{1\le i\le t-1}\sigma_i)\\
&=(w-\lceil\frac{w}{t^2}\rceil)\frac{1}{t^2-t+1}+(1-\frac{1}{t^2-t+1})\sum_{1\le i\le t-1}\sigma_i\\
&=(w-\lceil\frac{w}{t^2}\rceil)\frac{1}{t^2-t+1}+(1-\frac{1}{t^2-t+1})\sum_{1\le i\le t-2}\sigma_i+(1-\frac{1}{t^2-t+1})\sigma_{t-1}\\
&\ge (w-\lceil\frac{w}{t^2}\rceil)[\frac{1}{t^2-t+1}+(1-\frac{1}{t^2-t+1})\frac{1}{t^2-(t-1)+1}]\\
&\quad\ +(1-\frac{1}{t^2-t+1})(1-\frac{1}{t^2-(t-1)+1})\sum_{1\le i\le t-2}\sigma_i\\
&\ge (w-\lceil\frac{w}{t^2}\rceil)[\frac{1}{t^2-t+1}+(1-\frac{1}{t^2-t+1})\frac{1}{t^2-(t-1)+1}\\
&\quad\ +(1-\frac{1}{t^2-t+1})(1-\frac{1}{t^2-(t-1)+1})\frac{1}{t^2-(t-2)+1}]\\
&\quad\ +(1-\frac{1}{t^2-t+1})(1-\frac{1}{t^2-(t-1)+1})(1-\frac{1}{t^2-(t-2)+1})\sum_{1\le i\le t-3}\sigma_i\\
&\ge \cdots\\
&\ge (w-\lceil\frac{w}{t^2}\rceil)\sum_{2\le i\le t}\frac{1}{t^2-i+1}\prod_{i+1\le j\le t}(1-\frac{1}{t^2-j+1})+\prod_{2\le i\le t}(1-\frac{1}{t^2-i+1})\sigma_1\\
&= (w-\lceil\frac{w}{t^2}\rceil)\sum_{2\le i\le t}\frac{1}{t^2-i+1}\cdot\frac{t^2-t}{t^2-i}+\frac{t}{t+1}\sigma_1\\
&= (w-\lceil\frac{w}{t^2}\rceil)(t^2-t)\sum_{2\le i\le t}(\frac{1}{t^2-i}-\frac{1}{t^2-i+1})+\frac{t}{t+1}\sigma_1\\
&=(w-\lceil\frac{w}{t^2}\rceil) \frac{1}{t+1}+\frac{t}{t+1}\sigma_1.
\end{split}
\end{equation*}
Hence
\begin{equation*}
\begin{split}
\sum_{2\le i\le t}\sigma_i &=\sum_{1\le i\le t}\sigma_i-\sigma_1\\
& \ge (w-\lceil\frac{w}{t^2}\rceil) \frac{1}{t+1}+\frac{t}{t+1}\sigma_1-\sigma_1\\
& =\frac{1}{t+1}(w-\lceil\frac{w}{t^2}\rceil-\sigma_1),
\end{split}
\end{equation*}
that is, inequality $(1)$ holds.
\qed

\end{document}